\documentclass[a4paper,11pt]{article}
\usepackage{subfigure}
\usepackage{amsfonts}
\usepackage[dvips]{color,graphicx}
\usepackage{graphicx,amssymb,amsmath,bm,latexsym}
\usepackage{stmaryrd}
\usepackage{bm}
\usepackage{CJK}
\usepackage{indentfirst}
\usepackage{amsmath}

\newcommand{\reffig}[1]{Fig. \ref{#1}}

\definecolor{red  }{rgb}{1,0,0}
\definecolor{blue }{rgb}{0,0,1}
\definecolor{green}{rgb}{0,1,0}

\DeclareMathOperator{\sech}{sech}

\usepackage{graphicx,amssymb,amsmath,bm,latexsym}

\begin{document}
\title{Soliton molecules  in  Sharma-Tasso-Olver-Burgers equation}
\author{Zhaowen Yan$^1$ and Senyue Lou$^2$\thanks{E-mail:lousenyue@nbu.edu.cn}\\
\footnotesize $^1$\it School of Mathematical Sciences, Inner Mongolia University, Hohhot, 010021, China\\
\footnotesize $^2$\it School of Physical Science and Technology, Ningbo University, Ningbo, 315211, China}
\date{\today}
\maketitle

\begin{abstract}
Soliton molecules have been experimentally discovered in optics and theoretically investigated for coupled systems. This paper is concerned with the formation of soliton molecules by the resonant mechanism for a noncoupled system, the Sharma-Tasso-Olver-Burgers (STOB) equation. In terms of introducing velocity resonance conditions, we derive the soliton (kink) molecules, half periodic kink (HPK) molecules and breathing soliton molecule of STOB equation. Meanwhile, the fission and fusion phenomena among kinks, kink molecules, HPKs and HPK molecules have been revealed. Moreover, we also discuss the central periodic kink solutions from the multiple solitary wave solutions.
\end{abstract}

\section{Introduction}

The Burgers system and the Sharma-Tasso-Olver (STO) equation are  widely investigated from both mathematical and physical point of view \cite{Olver}-\cite{Liu15}. It is known that STO equation is an odd ordered equation of the Burgers hierarchy which has extensive application in  physical and engineering fields, such as the plasma physics,  fluid mechanics and statistical physics.
Lots of approaches have been applied to study STO equation including first integral method\cite{Lu12}, bi-Hamiltonian formulation method \cite{Tasso}, the fractional sub-equation method \cite{Zhang11},  the sine-cosine method \cite{WAM04}  and $q$-functions approach \cite{VVG97}. Furthermore, several transformations which involve Cole-Hopf transformation, fractional complex transform \cite{HJH12}, Darboux transformation \cite{Chen10} and B\"{a}cklund transformation have been used to discuss the STO equation \cite{Chen10,WTL}.

Solitons have been experimentally found in plasma physics and optics, as well as in nonlinear science area including Bose-Einstein condensation and DNA mechanical waves \cite{DP,L16}. It is well-known that solitons have an important application in a variety of areas, such as atmospheric and ocean dynamics \cite{Lou17}, optical fibers \cite{YN89}-\cite{MMC14}, photonic crystals \cite{SS99} and plasmas \cite{L83}. Later on the soliton molecules which are the soliton bound states have attracted considerable attention. Soliton molecules have been observed in optical systems \cite{SP05}-\cite{Nano} and analyzed in  Bose-Einstein condensates \cite{LNS}.  Various theoretical proposals to form soliton molecules have been established \cite{CK03}-\cite{KS11}. It is known that soliton molecules in coupled systems have been well discussed \cite{SK97}.  Recently, the breathing soliton molecules have been experimentally observed in a mode-locked fiber laser \cite{SA}.

The resonant theory of solitons  is applicable to a variety of integrable systems, such as  the KP-(II) equation \cite{YK10,KY16} and Novikov-Veselov equation \cite{JC} in which Wronskian representation of the $\tau$-function  have been employed. By restricting different resonant conditions on the solitons, it may derive many types resonant excitations. The resonant solutions have been generated by using linear superposition principle and the Bell polynomials for the bilinear equations \cite{MWX}. By virtue of Hirota¡¯s direct method and the B\"{a}cklund transformation, the researchers study the fission and fusion of the solitary wave solution of the Burgers and STO equations \cite{WTL}.  Furthermore,  the interaction between the lump soliton and the pair of resonance stripe solitons form a rogue wave solution \cite{CT18}. The rational solutions to the modified Korteweg-de Vries equation have been analyzed by imposing  the wave number resonance constraints \cite{ZDJ14}. Lately, by means of introducing velocity resonant mechanism, Lou investigated some types of soliton molecules in fluid systems \cite{Lou19} and nonlinear optical systems \cite{XL} involving the third-fifth order Korteweg de-Vries (KdV) equation, the KdV-Sawada-Kotera (KdVSK) equation, the KdV-Kaup-Kupershmidt (KdVKK) equation, the Hirota system and the potential modified KdV-sG system which derive kink-kink molecule, kink-antikink molecule, kink-breather molecule and breather-breather molecule etc.
Although the structure and properties of the integrable systems have been widely investigated, much less is  known about soliton molecules of the integrable systems. Our objective in this work is to  build new types of  molecules in a noncoupled system, i.e., the STOB equation by using velocity resonant conditions.

The article is structured as follows. Sec. II introduces the fission and fusion phenomenon of soliton solutions  for the STOB equation. In Sec. III,  we present some properties including the fission and fusion among the half periodic kink (HPK) and kink solutions.  We also develop  central periodic kink solutions for STOB equation. The fission and fusion among the soliton molecules, HPK molecules and kink molecules have been studied in Sec. IV.  In addition, we  derive a dissipative soliton molecule of multiple solitary solution for STOB equation.  Section IV is devoted to summary and conclusion.

\section{The fission  and fusion in Sharma-Tasso-Olver-Burgers equation}

Firstly, we give a brief introduction of the Sharma-Tasso-Olver-Burgers (STOB) equation  which will be useful in what follows. We consider the nonlinear evolution equation STOB:
\begin{eqnarray}\label{STOB}
u_t+\alpha(3 u_x^2+3u^2u_x+3 u u_{xx}+ u_{xxx})+\beta(2uu_x+u_{xx})=0,
\end{eqnarray}
for which $\alpha=0$  reduces to the Burgers equation \cite{BurgersE}, while $\beta=0$ represents the STO equation \cite{STO97}.

With the aim of construction of the solutions for STOB equation, we introduce the truncated Painlev\'{e} expansion of the STOB equation
\begin{eqnarray}\label{PE}
u=\sum_{j=0}^\alpha u_j v^{j-\alpha},
\end{eqnarray}
where $u_j$ are  functions of the derivatives of $v$. Especially, substituting $\alpha=1$ into (\ref{PE}), we obtain
\begin{eqnarray}\label{PE1}
u=\frac{u_0}{v}+u_1,
\end{eqnarray}
where $u_1$ is an arbitrary solution of Eq.(\ref{STOB}). Eq.(\ref{PE1}) is a B\"{a}cklund transformation of Eq.(\ref{STOB}). By substituting (\ref{PE1}) with $u_1=0$  into Eq.(\ref{STOB}), we derive a equation involves the relation between $u_0$ and $v$. Restricting the coefficient of $v^{-4}$ being vanished, we have
\begin{eqnarray}\label{v4}
3\alpha v^2 u_0^2 -\alpha u_0^3 v_x-2\alpha u_0^3 v_x^3=0.
\end{eqnarray}
Solving the Eq.(\ref{v4}), we get
\begin{eqnarray}\label{uv}
u_0=v_x,\ \ \ u_0=2v_x.
\end{eqnarray}
Here we only consider the first case. Inserting  $u_0=v_x,u_1=0$ into (\ref{PE1}), we have the following Cole-Hopf transformation
\begin{eqnarray}\label{truv}
u=\frac{v_x}{v}.
\end{eqnarray}
If we plug (\ref{truv}) back into (\ref{STOB}) and solve the solution, we obtain
\begin{eqnarray}\label{Solu}
v_t=-\alpha v_{xxx}-\beta v_{xx}-c v,
\end{eqnarray}
where $c$ is a function of $t$. Without loss of generality, we can simply take $c=0$.

Firstly, let us concentrate on a multiple solitary wave solution which possesses the form
\begin{eqnarray}\label{a3}
v=1+\sum_{i=1}^{N}a_ie^{\omega_{i}t+k_i x+\xi_{i}}
\end{eqnarray}
with the wave numbers $k_i$, frequencies $\omega_i$ and original positions $\xi_i$. In \eqref{a3}, $a_i$ have been inserted for convenience though they can be taken as $1$ without loss of generality.

It is straightforward to show that (\ref{a3}) is a solution of the STOB equation under the dispersion relations
\begin{eqnarray}\label{w}
\omega_i=-\alpha k_i^3-\beta k_i^2, \ \ i=1, \ldots,\ N.
\end{eqnarray}
Inserting (\ref{a3}) with $N=2, a_i=1$ and (\ref{w}) into (\ref{truv}), we obtain the two solitary wave solution
\begin{eqnarray}\label{u2}
u=\frac{k_1e^{-\alpha k_1^3 t-\beta k_1^2 t+k_1 x+\xi_{1}}+k_2e^{-\alpha k_2^3 t-\beta k_2^2 t+k_2 x+\xi_{2}}}{1+e^{-\alpha k_1^3t-\beta k_1^2 t+k_1 x+\xi_{1}}+e^{-\alpha k_2^3 t-\beta k_2^2 t+k_2 x+\xi_{2}}}.
\end{eqnarray}
It is easy to obtain the potential field $z=u_x$ of (\ref{u2})
\begin{eqnarray}\label{filed2}
z&=&\frac{k_1^2e^{-\alpha k_1^3 t-\beta k_1^2 t+k_1 x+\xi_{1}}+k_2^2 e^{-\alpha k_2^3 t-\beta k_2^2 t+k_2 x+\xi_{2}}}{1+e^{-\alpha k_1^3 t-\beta k_1^2 t+k_1 x+\xi_{1}}
+e^{-\alpha k_2^3 t-\beta k_2^2 t+k_2 x+\xi_{2}}}\\ \notag
&&-\frac{(k_1 e^{-\alpha k_1^3 t-\beta k_1^2 t+k_1 x+\xi_{1}}
+k_2 e^{-\alpha k_2^3 t-\beta k_2^2 t+k_2 x+\xi_{2}})^2}{(1
+e^{-\alpha k_1^3 t-\beta k_1^2 t+k_1 x+\xi_{1}}
+e^{-\alpha k_2^3 t-\beta k_2^2 t+k_2 x+i\xi_{2}})^2}.
\end{eqnarray}

Let us concentrate on the fission and fusion phenomenon for the two solitary wave solutions (\ref{u2}).
Soliton fission is the splitting of  a soliton to produce two or more solitons.  More then two solitons generate a soliton which is called soliton fusion. Let us present the limit expression for the potential field $z$.
For the soliton fusion case \\
(a) For $k_1k_2<0, $
\begin{eqnarray}\label{fu1}
z\rightarrow
\begin{cases}
  z_1+z_2,\ \ \ t\rightarrow -\infty,\\
  z_3,  \ \ \ \ \ \ \ \ \ \  t\rightarrow +\infty,
\end{cases}
\end{eqnarray}
where
\begin{eqnarray}\label{z123}
z_1&=&\frac{k_1^2}{4} \sech^2\bigg[\frac{1}{2}k_1(x-(\alpha k_1^2+\beta k_1)t+\xi_{1})\bigg],\label{so1}\\
z_2&=&\frac{k_1^2}{4} \sech^2\bigg[\frac{1}{2}k_2(x-(\alpha k_2^2+\beta k_2)t+\xi_{2})\bigg],\label{so2}\\
z_3&=&\frac{(k_1-k_2)^2}{4} \sech^2\bigg[\frac{1}{2}(k_1-k_2)[x-(\alpha( k_1^2+k_2^2+k_1k_2)+\beta( k_1+k_2))t ]\notag\\
&&+\xi_{1}-\xi_{2}\bigg]. \label{so3}
\end{eqnarray}\\
(b) For $k_1k_2>0, $
\begin{eqnarray}\label{fu2}
z\rightarrow\begin{cases}
  z_2+z_3,\ \  t\rightarrow -\infty,\\
  z_1,  \ \ \ \ \  \ \  \ \ t\rightarrow +\infty.
\end{cases}
\end{eqnarray}
Concentrating on soliton fission, we have\\
(c) For $k_1k_2<0, $
\begin{eqnarray}\label{fi1}
z\rightarrow\begin{cases}
  z_1,\ \ \ \ \ \ \ \ \ t\rightarrow -\infty,\\
  z_2+z_3,  \ \ \ t\rightarrow +\infty.
\end{cases}
\end{eqnarray}
(d) For $k_1k_2>0, $
\begin{eqnarray}\label{f2}
z\rightarrow\begin{cases}
z_3,\ \ \ \ \ \ \ \ \ \  t\rightarrow -\infty,\\
  z_1+z_2,  \ \ \ \ t\rightarrow +\infty.
\end{cases}
\end{eqnarray}

The single soliton solutions of  (\ref{so1})  and (\ref{so2}) with $k_1k_2<0$ are the exact solutions of STOB equation. Then two solitary waves occur interaction which makes two solitary waves fuse to a resonant solution by (\ref{so3}). Based on (\ref{fu1}), it is obvious that before the fusion, the two single solitary waves lie on $x=(\alpha k_1^2+\beta k_1)t-\xi_{1}$ and $x=(\alpha k_2^2+\beta k_2)t-\xi_{2}$ with amplitudes $\frac{1}{4}k_1^2$ and $\frac{1}{4}k_2^2$ respectively. While after fusion, the resonant solitary wave changes its location to $\alpha(( k_1^2+k_2^2+k_1k_2)+\beta( k_1+k_2))t-\xi_{1}+\xi_{2}$ with the amplitude $\frac{1}{4}(k_1-k_2)^2$.

\begin{figure}[htbp]
\centering
\subfigure(I){\includegraphics[height=5.5cm,width=6cm]{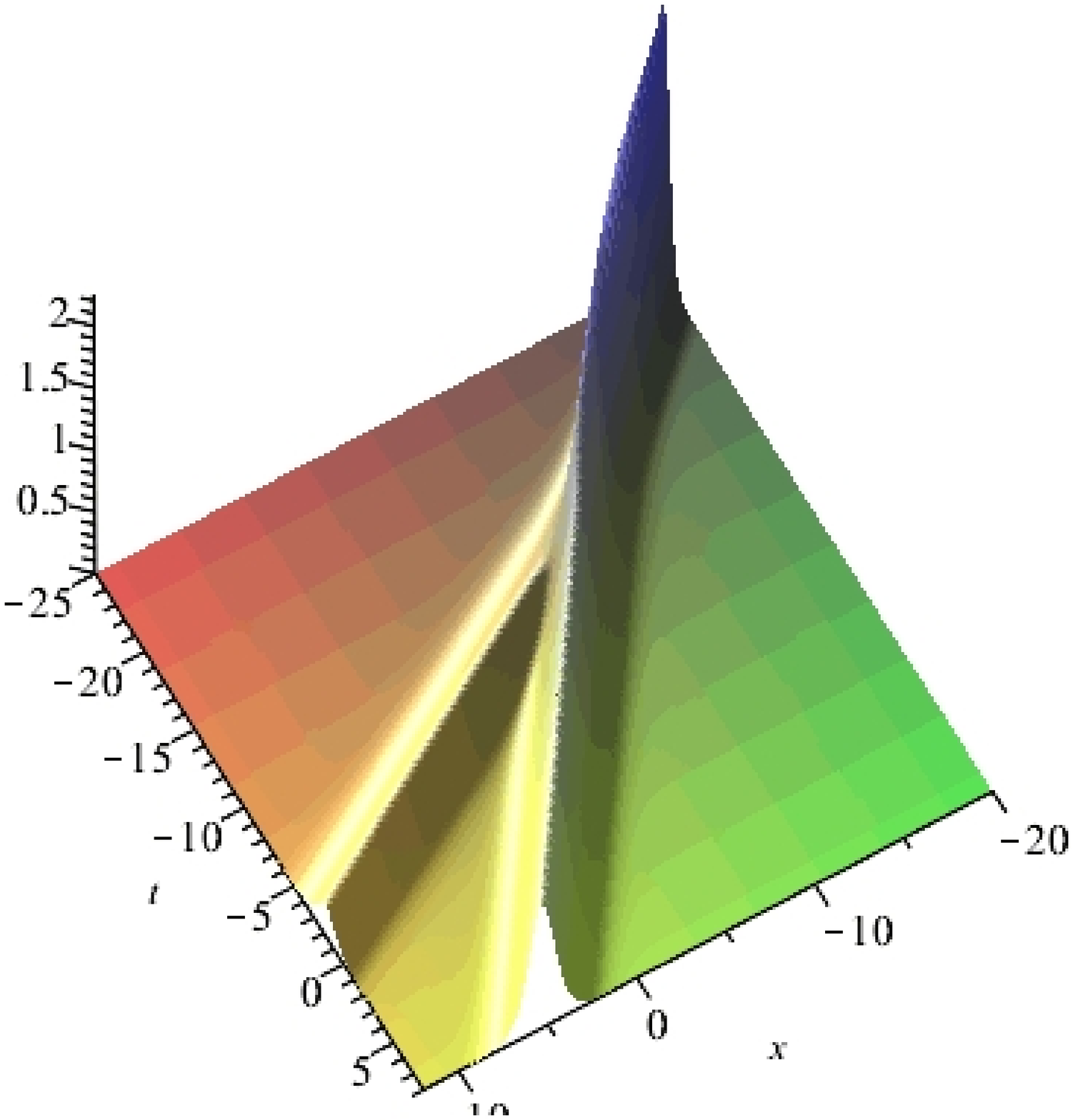}}
\subfigure(II){\includegraphics[height=5.5cm,width=6cm]{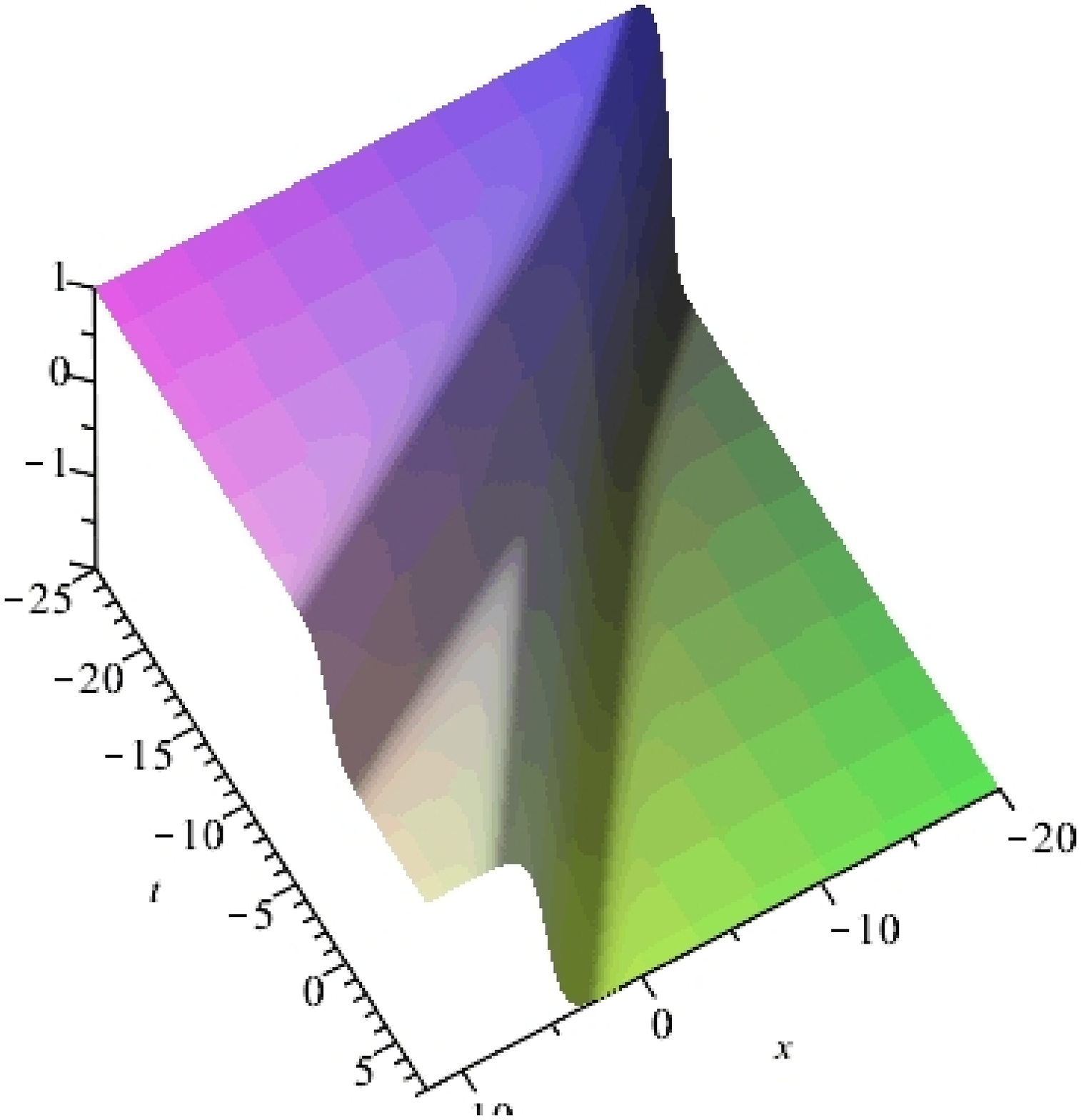}}
\caption{ Plot of two solitary wave fission of STOB  equation with the parameter
selections \eqref{ss2p}: (I) $z$; (II) $u$.}
\label{rufission}
\end{figure}

\reffig{rufission}  is a plot of the fission phenomenon of the two solitary waves with parameters given by
\begin{eqnarray}\label{ss2p}
k_1 = 1, \ k_2 = -2,\  \alpha = \frac{2}{3},\  \beta = 1, \ \xi_{1} = -15, \ \xi_{2} = 0.
\end{eqnarray}
It should be noted  that one resonant solitary wave fission  to two single solitary wave at one time.

\input

\begin{figure}[htbp]
\centering
\subfigure(I){\includegraphics[height=5.5cm,width=6cm]{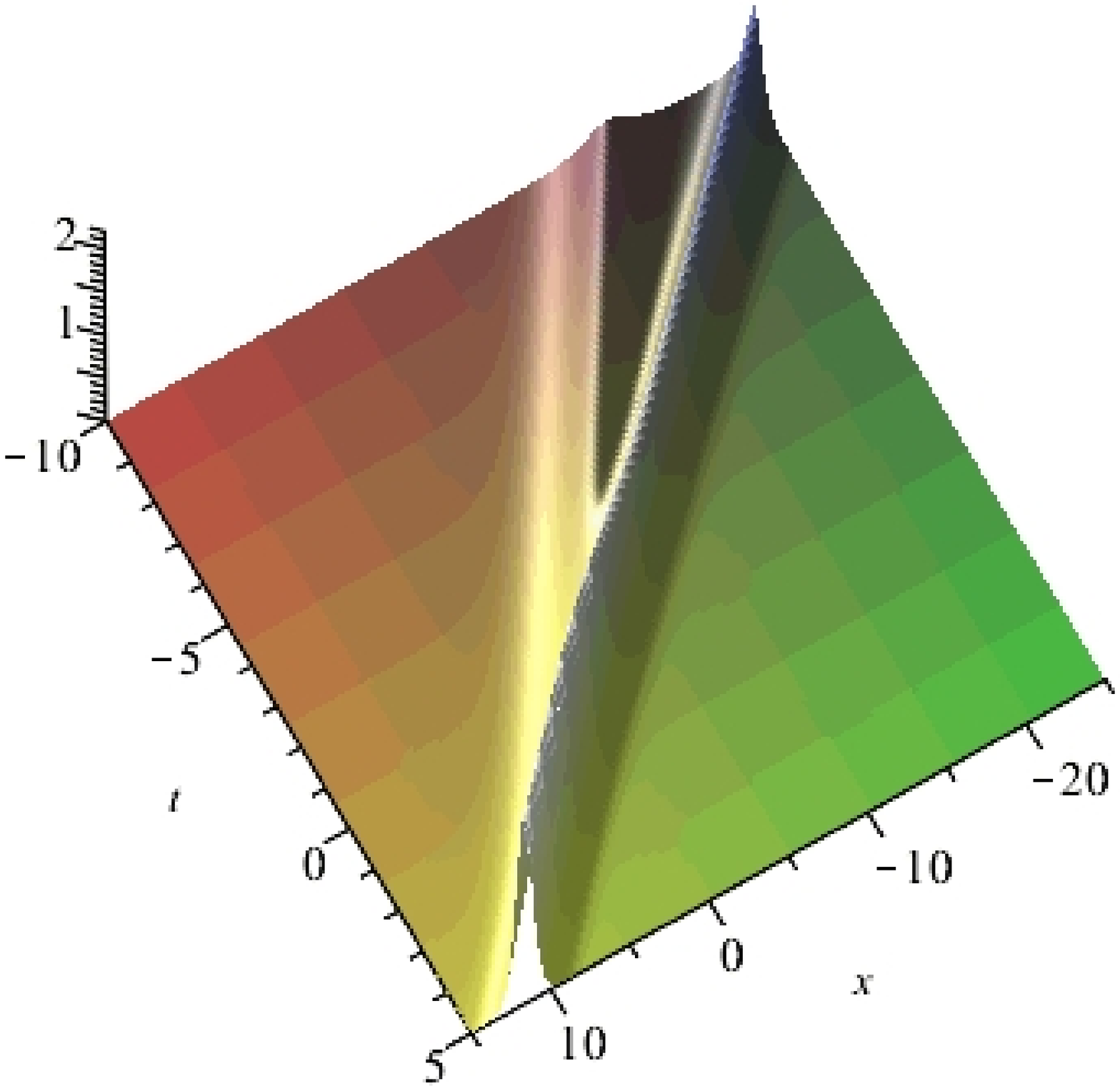}}
\subfigure(II){\includegraphics[height=5.5cm,width=6cm]{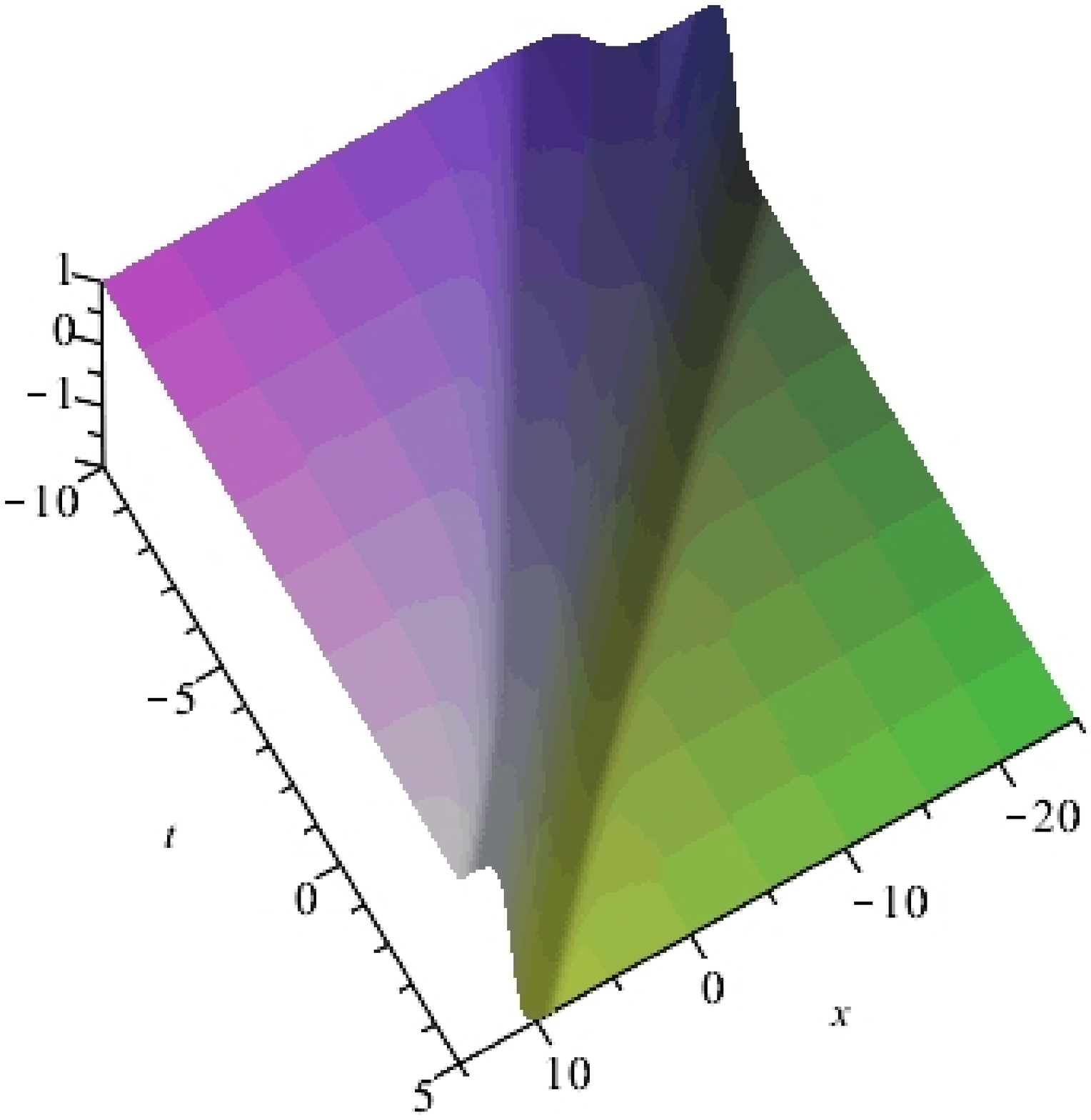}}
\caption{Plot of two solitary wave fusion of STOB  equation with the parameter
selections \eqref{ss3p}  (I) $z$; (II) $u$.}
\label{rufusion}
\end{figure}

\reffig{rufusion} is a plot of the fusion phenomenon of the two solitary waves with parameter selections
\begin{eqnarray}\label{ss3p}
k_1 = 1,\  k_2 = -2, \  \alpha =1, \ \beta = \frac{2}{3},\  \xi_{1} = 0,\  \xi_{2} = 0.
\end{eqnarray}
From \reffig{rufusion}, we find two single solitary waves fusion to one resonant solitary wave at some time.

Based on \reffig{rufission} and \reffig{rufusion}, we conclude the soliton's  velocity, amplitude, width and wave shape changed after the collision.  It is known that only fusion and no fission happen in the Burgers equation, while both fission and fusion occur for STO and STOB equations.

\section{The half periodic kink solutions and their fissions and fusions}

In this section, we turn to investigate some special structures and the properties of the solutions which are valid only for the STOB equation but not for the Burgers and STO equations.

 From the expression of the two soliton solution, we know that if the wave number $k_2$ and the position parameter $\xi_2$ are taken as the complex conjugates of $k_1=k+\mbox{i}\kappa$ and $\xi_1=\eta+\mbox{i}\zeta $, the solution will become a singular kink (or named singular complexiton),
\begin{eqnarray}\label{comp}
u=\frac{a\exp(k x-\omega t+\eta)[k\cos(\kappa x-\Omega t+\zeta)-\kappa \sin(\kappa x-\Omega t+\zeta)]}{1+a\exp(k x-\omega t+\eta)\cos(\kappa x-\Omega t+\zeta)}.
\end{eqnarray}
with $a=2a_2=2a_1,\ \omega=\alpha k (k^2-3\kappa^2)+\beta (k^2-\kappa^2),$ and $\Omega=[\alpha (3 k^2-\kappa^2)+2\beta k]\kappa$. The solution \eqref{comp} is singular at the points $1+a\exp(k x-\omega t+\eta)\cos(\kappa x-\Omega t+\zeta)=0$.

In order to find nonsingular kink with periodic function, one more soliton should be added with the resonant conditions
\begin{equation}
k_3=k,\ \omega_3=-\alpha k_3^3-\beta k_3^2=-\omega.\label{HPKr}
\end{equation}
The resonant condition \eqref{HPKr} leads to $k=-\frac{\beta}{3\alpha}$.
Thus, we get the following solution with both periodic and kink behaviors,
\begin{eqnarray}\label{Case1}
u=-\frac1{3\alpha}\frac{b\beta+a\beta\cos(\kappa x-\Omega t+\zeta)+3a\alpha\kappa \sin(\kappa x-\Omega t+\zeta)}{\exp\left[\frac{\beta x}{3\alpha}+\frac{2\beta^3t}{27\alpha^2}-\eta\right]+b+a\cos(\kappa x-\Omega t+\zeta)}
\end{eqnarray}
with arbitrary real constants $a,\ b,\ \kappa,\ \eta,\ \zeta$ and $\Omega=-\frac{\kappa}{3 \alpha}(3\alpha^2\kappa^2+\beta^2)$.

It is clear that the solution \eqref{Case1} is analytic for the parameter condition $b>|a|$. From the expression \eqref{Case1}, we know that the solution tends to a constant zero for the half plane $\frac{\beta x}{3\alpha}+\frac{2\beta^3t}{27\alpha^2}-\eta>0$ and the solution will tend to a periodic wave
$$
u\rightarrow -\frac1{3\alpha}\frac{b\beta+a\beta\cos(\kappa x-\Omega t+\zeta)+3a\alpha\kappa \sin(\kappa x-\Omega t+\zeta)}{b+a\cos(\kappa x-\Omega t+\zeta)},\ \frac{\beta x}{3\alpha}+\frac{2\beta^3t}{27\alpha^2}-\eta\rightarrow -\infty
$$
at the other half plane $\frac{\beta x}{3\alpha}+\frac{2\beta^3t}{27\alpha^2}-\eta<0$. Thus, for simplicity later, we call the solution \eqref{Case1} as the half periodic kink (HPK).

\begin{figure}[htbp]
\centering
\subfigure(I){\includegraphics[height=5cm,width=6.5cm]{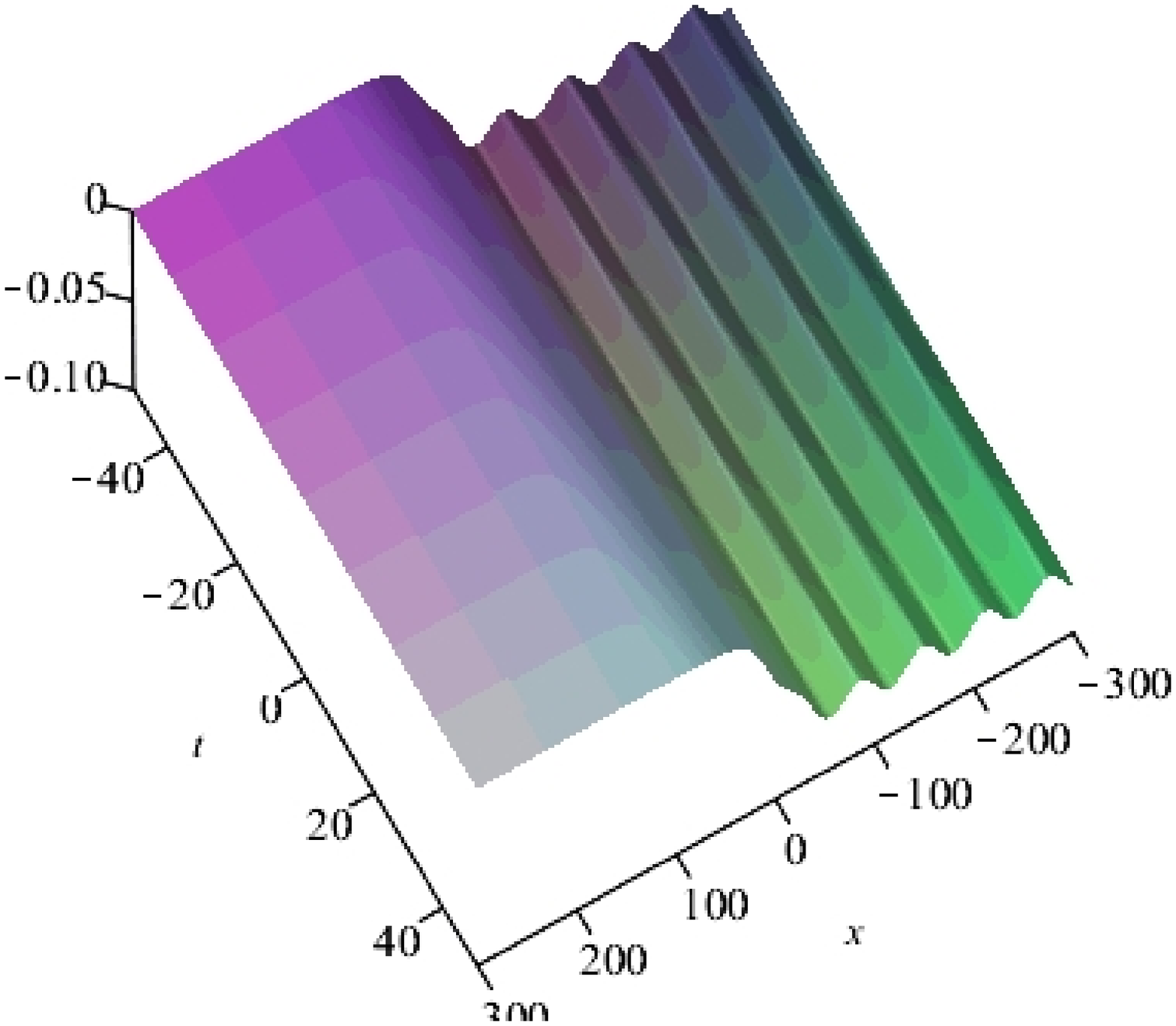}}
\subfigure(II){\includegraphics[height=5cm,width=6.5cm]{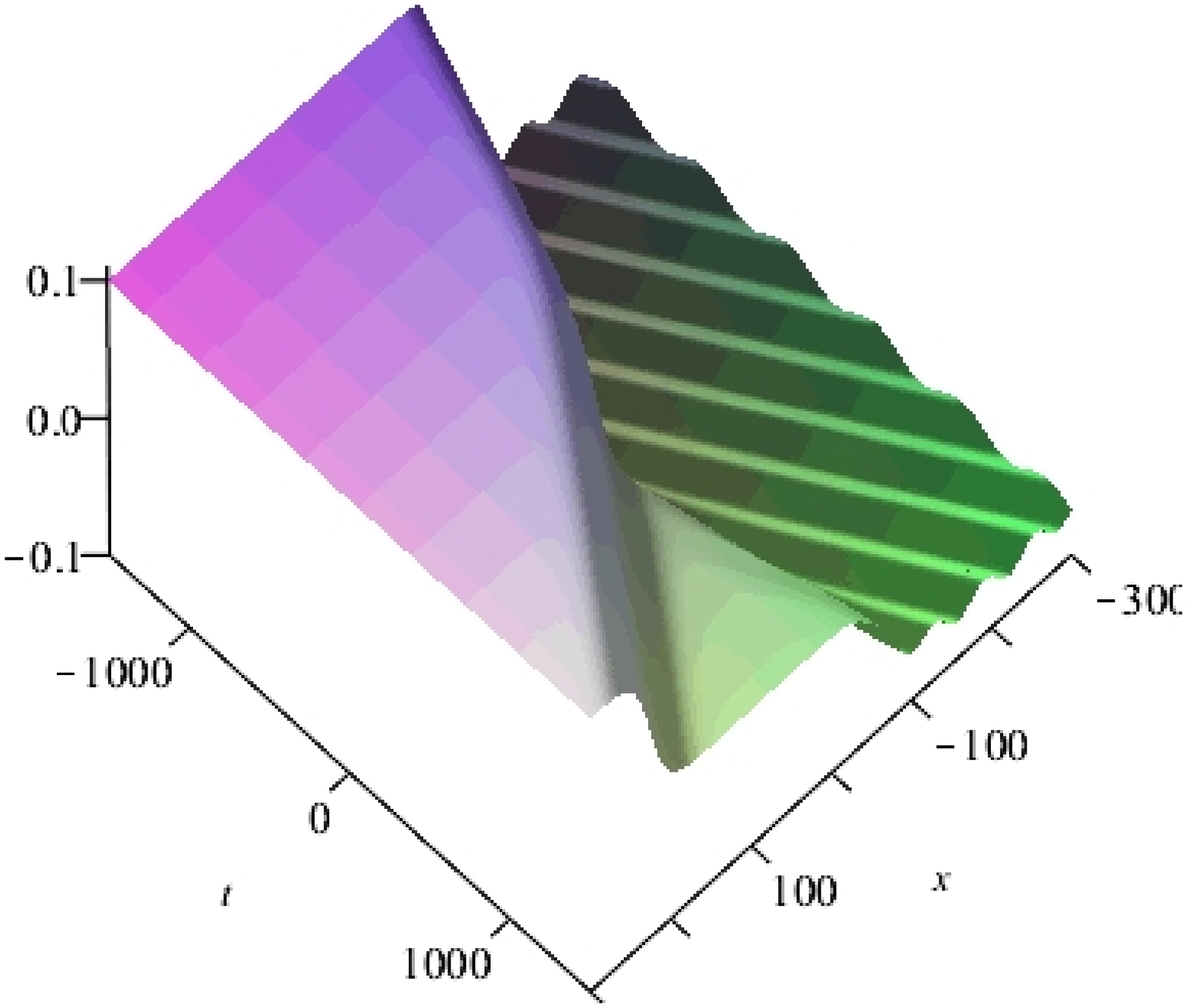}}
\caption{(I) HPK solution for STOB equation described by Eq. (\ref{Case1}) with the parameter selections \eqref{C1}, (II) The HPK fissions to one usual kink and one HPK described by Eq. (\ref{N}) with $N=4$ and the parameter selections \eqref{C2}.}\label{HPK1}
\end{figure}

\reffig{HPK1} (I) indicates a special HPK structure for the STOB equation described by Eq. (\ref{Case1}) with the parameter selections
\begin{eqnarray}\label{C1}
 &&\ a=-k=\frac{1}{15},\ \kappa=\frac{1}{10},\ \alpha=5,\ \beta=b=1,\  \eta=\zeta=0.
\end{eqnarray}

\input epsf

Now let us consider the  fission and fusion phenomenon between the HPK solution and kink solution by adding more solitons. For $N$ soliton solutions, we have
\begin{eqnarray}\label{N}
u=\frac{\sum\limits_{i=1}^N a_ik_ie^{k_i x-\alpha k_i^3 t-\beta k_i^2 t+\xi_{i}}}{1+{\sum\limits_{i=1}^Na_ie^{k_i x-\alpha k_i^3 t-\beta k_i^2 t+\xi_{i}}}}.
\end{eqnarray}

A usual kink may be split out from an HPK.
\reffig{HPK1} (II) displays the HPK soliton fissions to an HPK and a usual kink for STOB equation described by (\ref{N}) with $N=4$ and the parameter selections
\begin{eqnarray}\label{C2}
 &&a_1= a_2=1,\ a_3=a_4=\frac{1}{30},\  k_1=\frac{1}{10}, \ k_2=-\frac{1}{15},\ k_3=-\frac{1}{15}+\frac{\mbox{i}}{10}, \notag\\
 &&\ k_4=-\frac{1}{15}-\frac{\mbox{i}}{10},\ \alpha=5,\ \beta=1,\ \xi_{j}=0,\ (j=1,\ldots, 4).
 \end{eqnarray}

\begin{figure}%ru81
\centering\epsfxsize=6.5cm\epsfysize=5.5cm\epsfbox{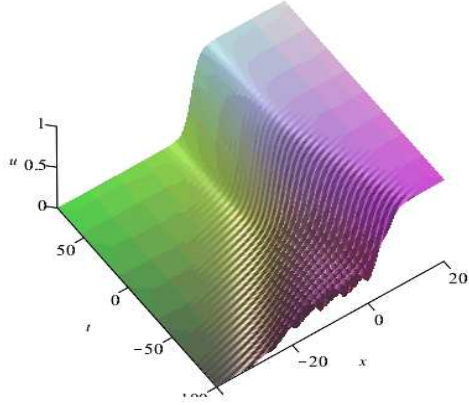}
\caption{Two HPKs are fused to an ordinary kink given by \eqref{N} with $N=4$ and  (\ref{para1}).} \label{HPKfusion}
\end{figure}

Two HPKs may be fused to one usual kink. \reffig{HPKfusion} demonstrates that two HPKs are fused to an ordinary kink described by \eqref{N} with $N=4$ and the parameter selections
\begin{eqnarray}\label{para1}
&&\ a_1=100,\ a_2=1,\ a_3=a_4=\frac{1}{20}, \ k_1=1,\ k_2=\frac{1}{3}, \ k_3=\frac{1}{3}+\mbox{i},\notag\\
&&\ k_4=\frac{1}{3}-\mbox{i}, \ \alpha=-1,\ \beta=1, \ \xi_{j}=0, (j=1,\ldots, 4).
\end{eqnarray}
From \reffig{HPK1} and \reffig{HPKfusion}, we have analyzed the fission and fusion between HPK and kink solutions. However, there
are no such solutions for the Burgers and STO equations

\input epsf

\section{Kink molecules and Soliton molecules for the STOB equation}

Soliton molecules are  bound states of solitons. In this section, we shall present some types of  molecules which contain soliton molecules, kink molecules, HPK molecules and breathing soliton moleculse  for STOB equation. The soliton molecules presented in this section exist only for the STOB equation but not the Burgers and STO equations.

To find possible soliton molecules for the STOB system \eqref{STOB}, we should introduce the velocity resonant conditions. If $i^{\mbox{th}}$ and $j^{\mbox{th}}$ solitons constitute a soliton molicule, then the velocity resonant condition
\begin{eqnarray}\label{RC}
\frac{\omega_i}{\omega_j}=\frac{k_i}{k_j},\ i\neq j, \ \ i,j=1,\ldots, N
\end{eqnarray}
should be satisfied. Using (\ref{w}) and (\ref{RC}),  the  velocity resonant condition \eqref{RC} becomes
\begin{eqnarray}\label{VRC}
k_j=-k_{i}-\frac{\beta}{\alpha}.
\end{eqnarray}

\begin{figure}[htbp]
\centering
\subfigure(I){\includegraphics[height=5cm,width=6.5cm]{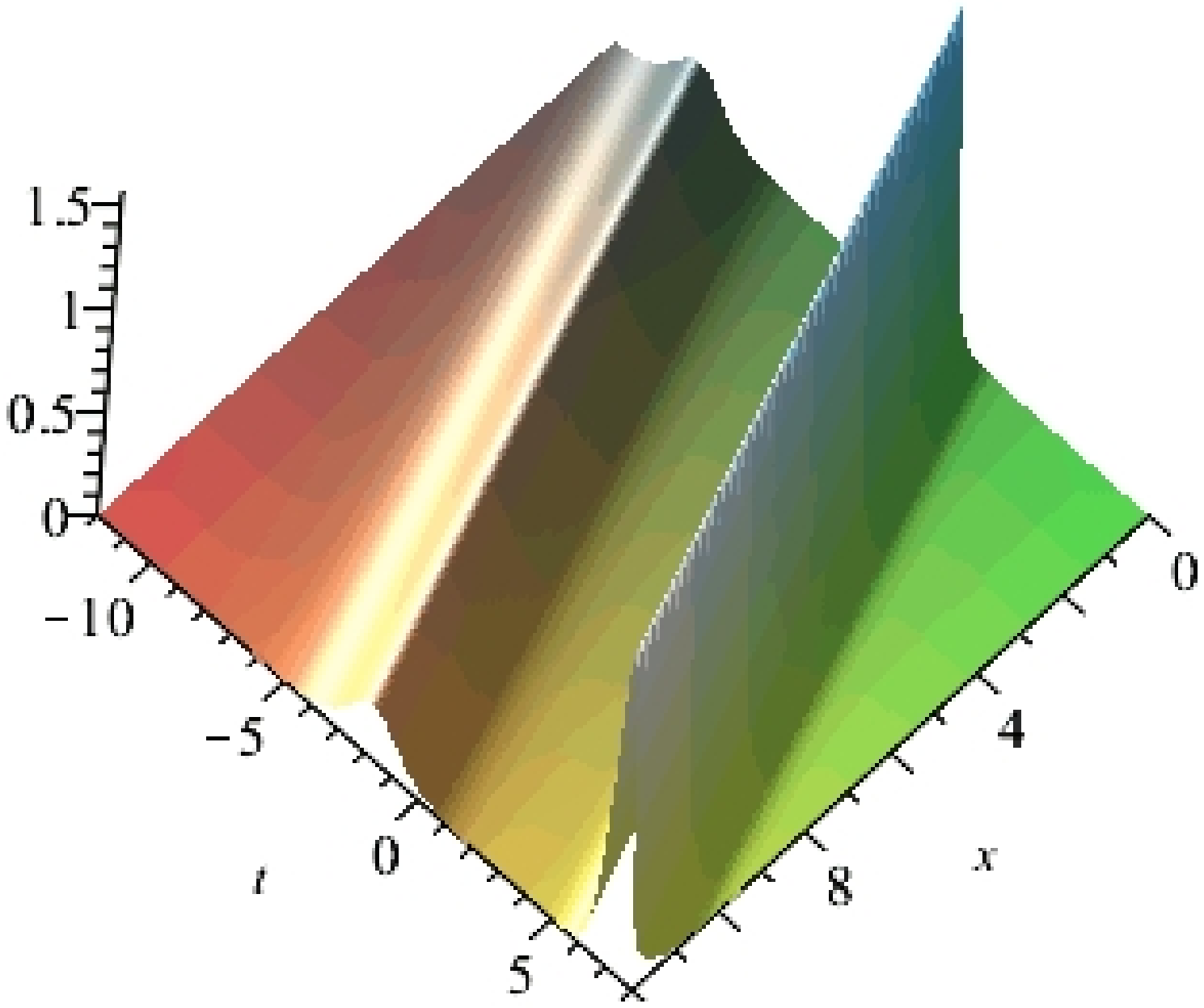}}
\subfigure(II){\includegraphics[height=5cm,width=6.5cm]{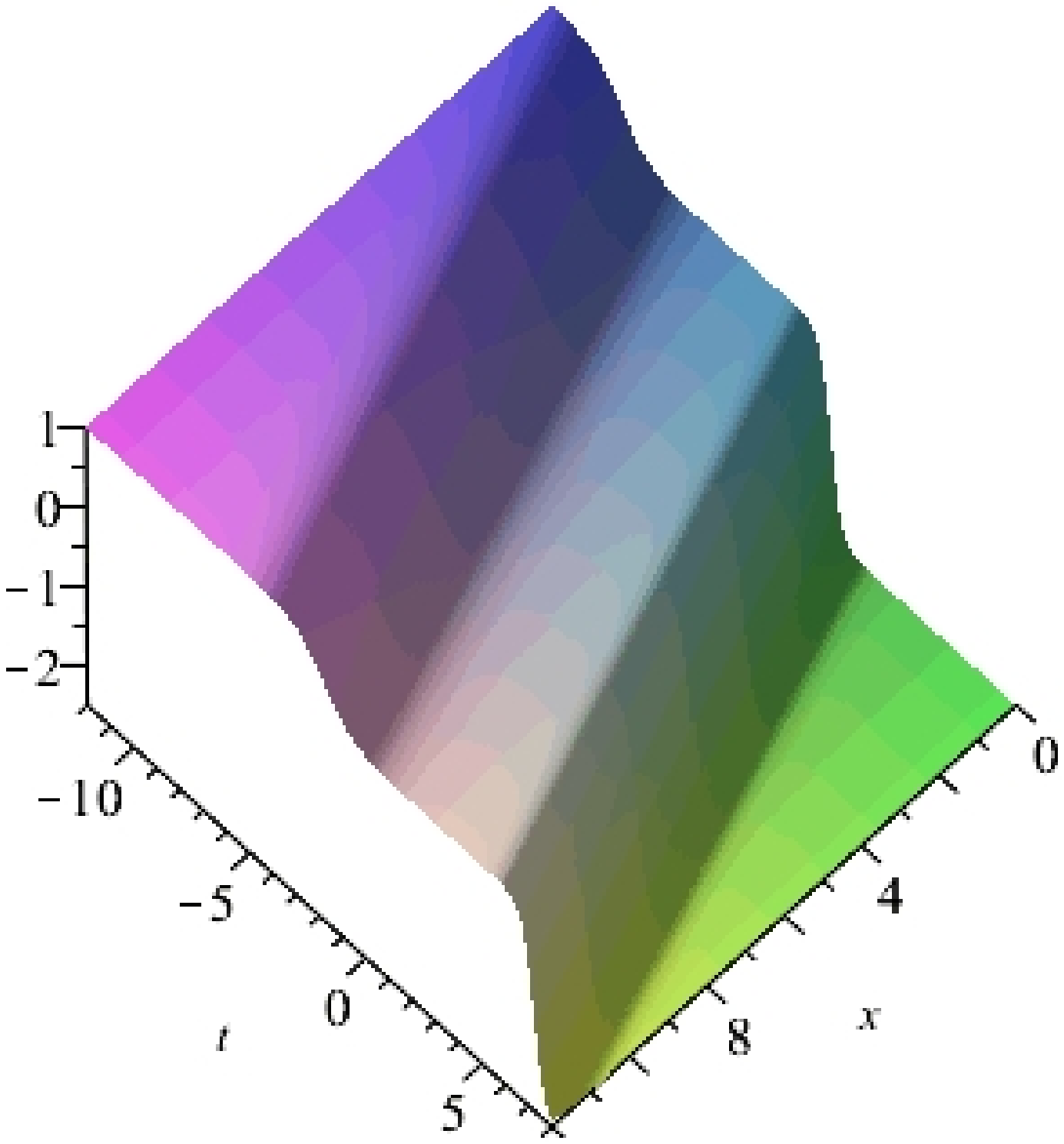}}
\caption{Soliton molecule structure for  STOB  equation expressed by (\ref{u2}) with the parameter
selections (\ref{PA1})   (I) $z$; (II) $u$.}\label{SM1}
\label{fig12d}
\end{figure}

From the resonant condition (\ref{RC}), the two solitary wave solutions are bounded to generate a soliton molecule. \reffig{SM1}  shows the plot of soliton molecule structure (\ref{u2}) of the STOB equation with the parameters being selected as
\begin{eqnarray}\label{PA1}
k_1=1, \ k_2=-\frac{5}{2},\   \alpha=\frac{2}{3},\ \beta=1,\ \xi_{1}=-15,\  \xi_{2}=0.
\end{eqnarray}
\input epsf
By using the velocity condition \eqref{VRC} for some pairs of solitons for $N$ soliton solutions, one may find fission and fusion phenomena among soliton molecules and solitons. For a multiple solitary wave solution (\ref{N}) with $N=4$, let us select the parameters as follows
\begin{eqnarray}\label{rsm2p}
 &&\alpha=5, \ \beta=1,\  k_1=\frac{1}{10},\ k_2=-\frac{1}{10}, \ k_3=-\frac{3}{10},\ k_4=-\frac{1}{10},\notag\\
 && \xi_{1}=-10, \ \xi_{2}=10, \ \xi_{3}=\xi_{4}=0, \ a_j=1, \ (j=1,\ldots,4).
\end{eqnarray}
Under the conditions (\ref{rsm2p}), we have
\begin{eqnarray}\label{u3}
u=\frac{\frac{1}{10}\mbox{\rm e}^{-\frac{3}{200}t+\frac{1}{10}x-10}
-\frac{1}{10}\mbox{\rm e}^{-\frac{1}{200}t-\frac{1}{10}x+10}
-\frac{3}{10}\mbox{\rm e}^{\frac{9}{200}t-\frac{3}{10}x}
-\frac{1}{10}\mbox{\rm e}^{-\frac{1}{200}t-\frac{1}{10}x}}{1
+\mbox{\rm e}^{-\frac{3}{200}t+\frac{1}{10}x-10}
+\mbox{\rm e}^{-\frac{1}{200}t-\frac{1}{10}x+10}
+\mbox{\rm e}^{\frac{9}{200}t-\frac{3}{10}x}
+\mbox{\rm e}^{-\frac{1}{200}t-\frac{1}{10}x}}.
\end{eqnarray}

\begin{figure}[htbp]
\centering
\subfigure(I){\includegraphics[height=5cm,width=6.5cm]{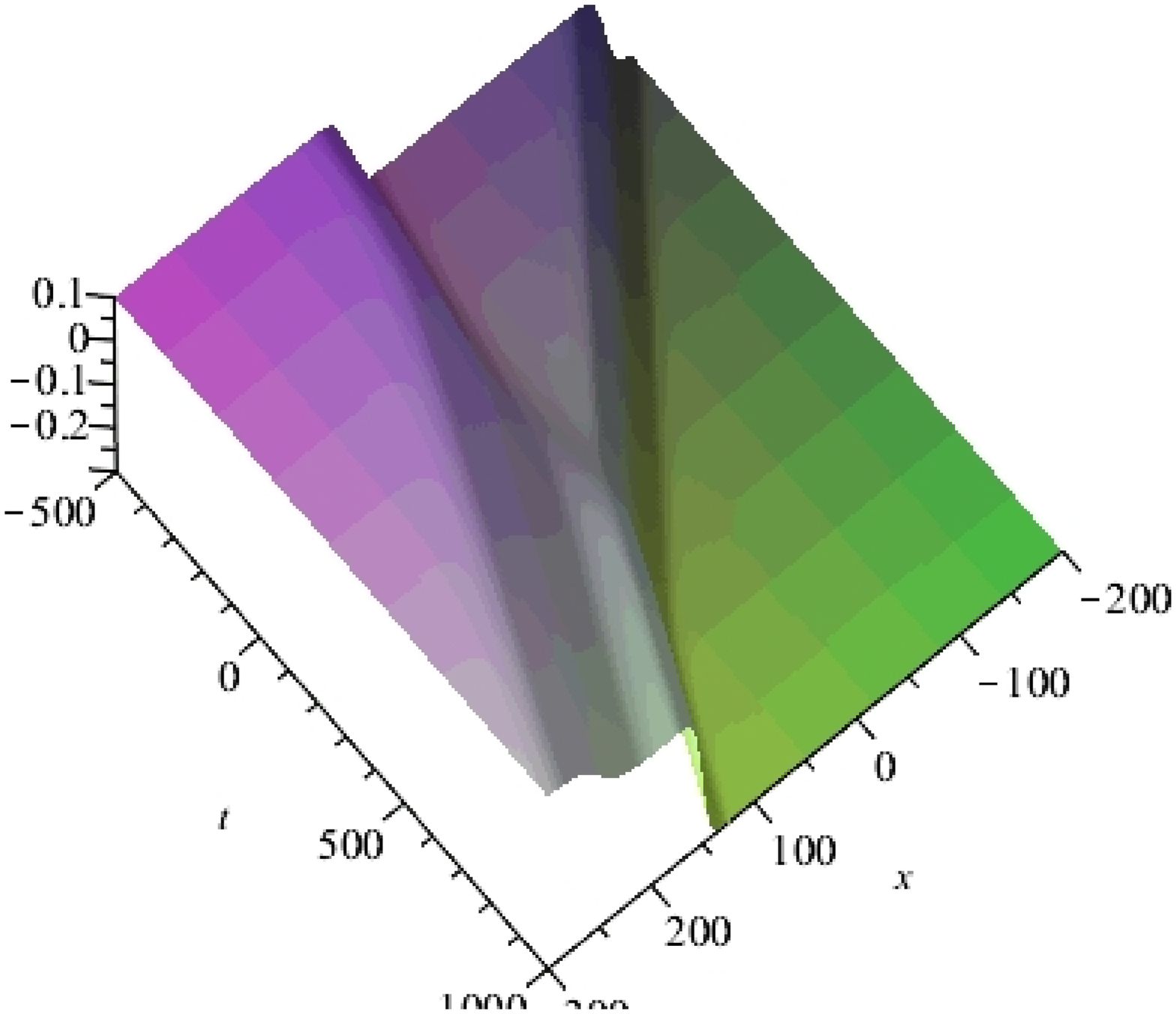}}
\subfigure(II){\includegraphics[height=5cm,width=6.5cm]{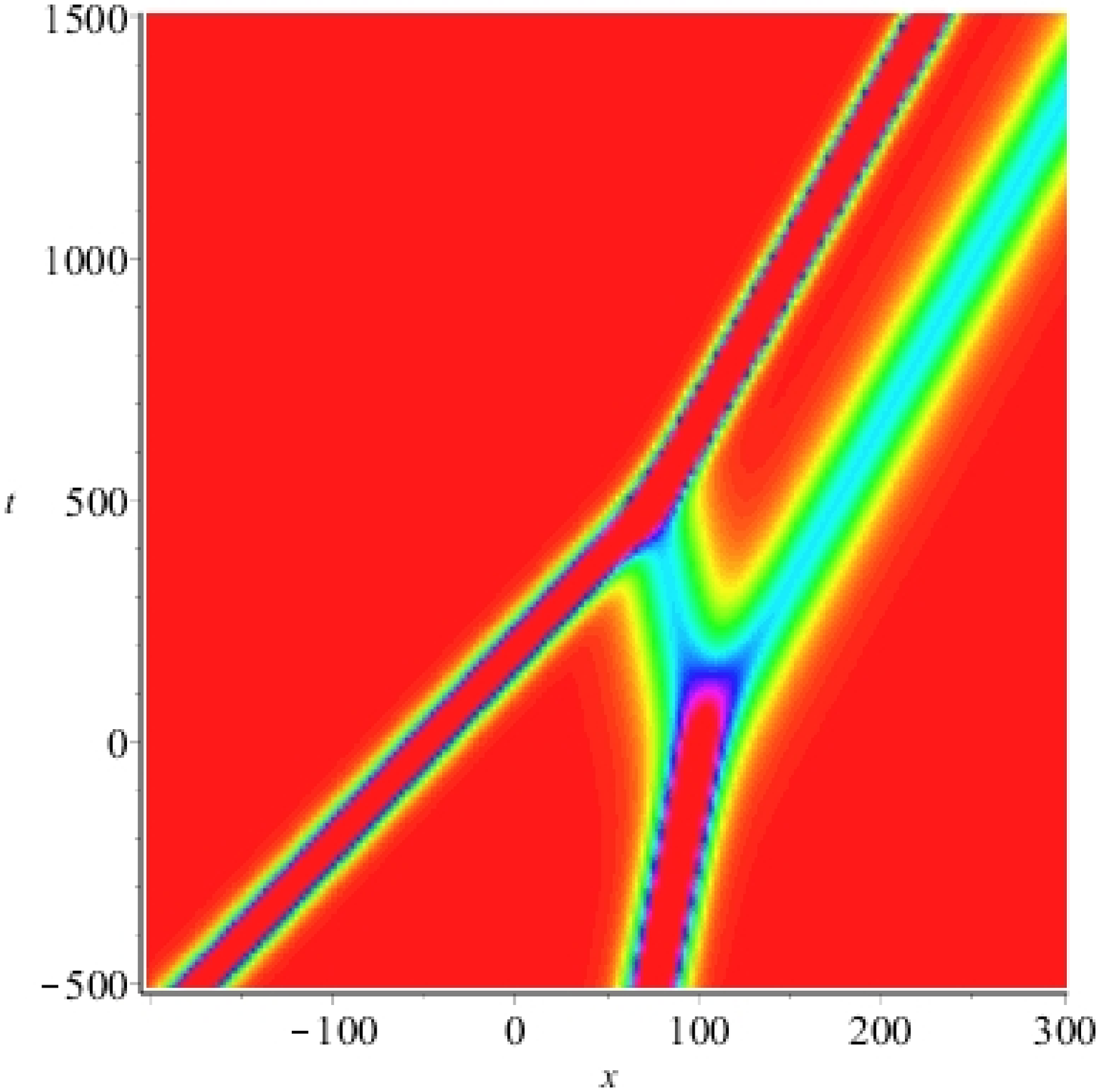}}
\caption{(I) Two kinks are fused to one kink molecule for the field $u$ described by Eq. (\ref{u3}), (II) Density plot of the interaction process (two solitons are fused to one soliton molecule) for the field $z=u_x$ with $u$ being described by Eq. (\ref{u3}).}
\label{SM2}
\end{figure}

\reffig{SM2} (I) is the plot of the four soliton molecule structure (\ref{u3}) with the parameter selections \eqref{rsm2p}. \reffig{SM2} (II) indicates the density plot of the molecule structure $z=u_x$ with $u$ being given by (\ref{u3}).
From \reffig{SM2} (II), one finds that the more detailed soliton interactions among four solitons expressed by \eqref{u3} can be divided into two parts, a fission interaction is followed by a fusion process. In other words, one soliton (right soliton from the bottom of the figure) is firstly split to two solitons (named soliton A (left) and soliton B (right)).  Then the soliton A and another soliton (left soliton from the bottom of the figure) are fused to one soliton which is combined with the soliton B  and then they form a soliton molecule.

\input epsf

In the previous section, we present some interesting  structure of the multiple solitary solutions of the STOB equation with paired conjugate complex parameters. Especially, the special new type of kink, HPK, solutions are found. Next we establish some different types of molecules from the multiple solitary wave solutions \eqref{N} by selecting the paired conjugate complex parameters.

For one paired complex conjugate wave numbers, we derive one simplest HPK molecule in the form $\left(\zeta=-\frac{2\beta^3}{27 \alpha^2} t-\frac{\beta}{3\alpha}x,\ \eta = -\kappa x -\frac{\kappa (3\alpha^2\kappa^2+\beta^2)}{3\alpha}t\right)$
\begin{eqnarray}\label{HPKM}
u&=&\frac1{3\alpha}\frac{6a_3\alpha\kappa\sin\eta-2a_3\beta\cos\eta -2 a_1\beta\exp\zeta-a_2\beta}
{2 a_3\cos\eta+2\sqrt{a_1}\cosh(\zeta+\ln\sqrt{a_1})+a_2}
\end{eqnarray}
by fixing the parameters in \eqref{N} with $N=4$ as
\begin{eqnarray}\label{P}
a_4=a_3,\ k_1=-\frac{2\beta}{3\alpha},\ k_2=-\frac{\beta}{3\alpha},\ k_3 = k_2+\mbox{\rm i}\kappa,\ k_4 = k_2-\mbox{\rm i}\kappa,\ \xi_{j}=0, \ (j=1,\ldots, 4).
\end{eqnarray}
Substituting parameters
\begin{eqnarray}\label{P3}
&&\alpha=5,\ \beta=1,\ a_1=1,\ a_2=1002,\ a_3=20, \kappa=-\frac13,
\end{eqnarray}
into Eq. \eqref{HPKM}, we have
\begin{eqnarray}\label{2.35}
u&=&-\frac{1}{15}\frac{\exp\zeta+20\cos\eta+100\sin\eta+501}
{\cosh\zeta+20\cos\eta+501},
\end{eqnarray}
where $\zeta=-\frac{2}{675}t-\frac{1}{15}x,\ \eta=\frac{28}{135}t+\frac{1}{3}x$.\\

\input epsf
\begin{figure}[htbp]
\centering
\subfigure(I){\includegraphics[height=5cm,width=6.5cm]{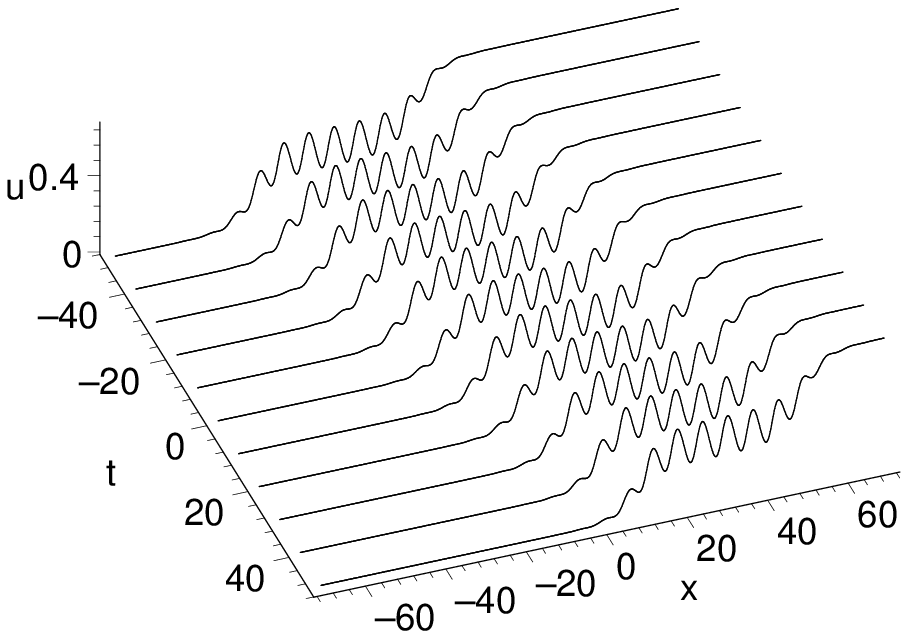}}
\subfigure(II){\includegraphics[height=5cm,width=6.5cm]{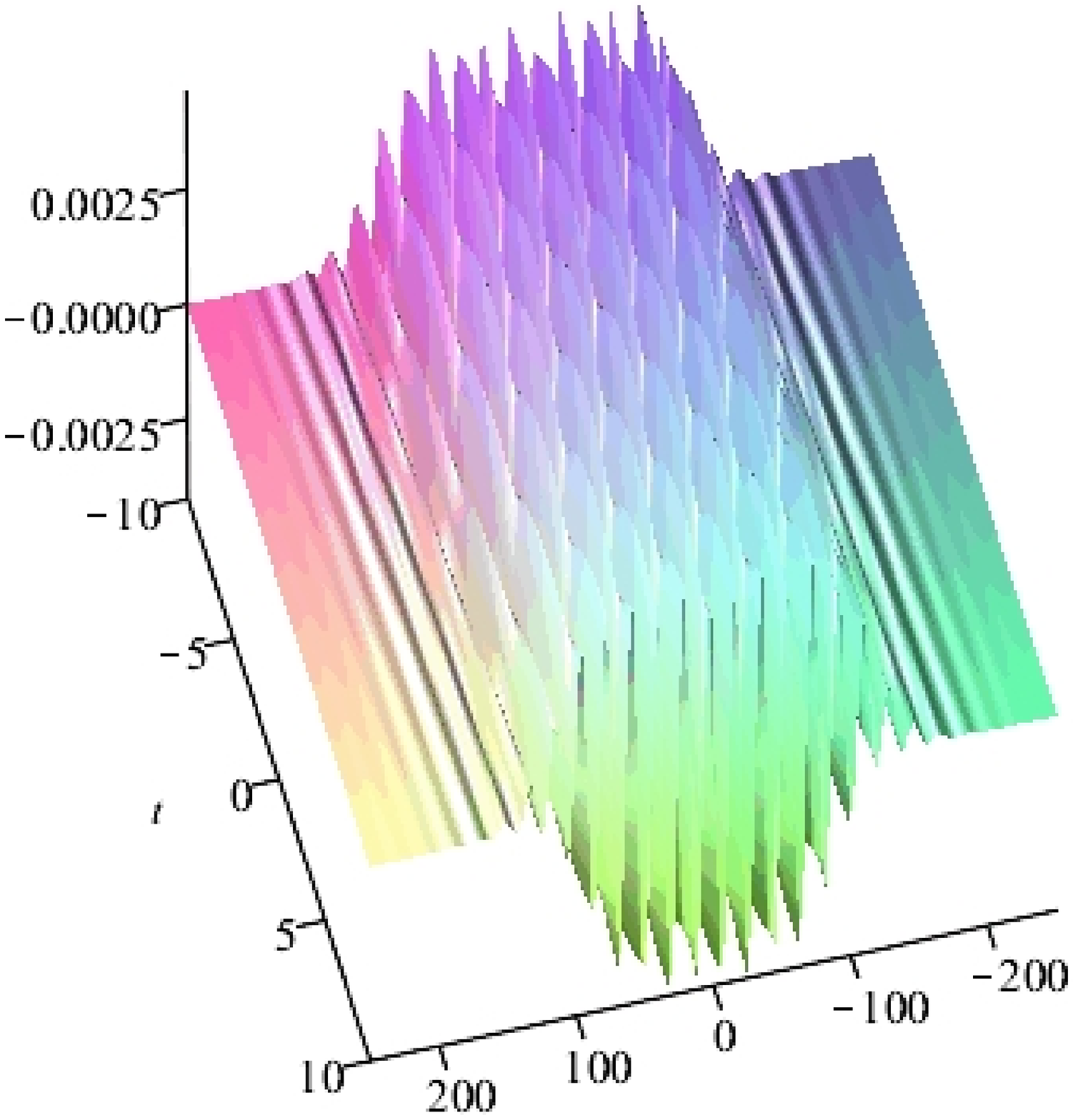}}
\caption{(I) The structure of the HPK  molecules described by  (\ref{2.35}), (II) Breathing dissipative solutions for $z$ of (\ref{2.35}).}\label{SM6}
\label{fig12d}
\end{figure}

\reffig{SM6} (I) shows  the HPK  molecules determined by  (\ref{2.35}). \reffig{SM6} (II) is the breathing dissipative solutions for the field $z=u_x$ while $u$ being given by (\ref{2.35}). The breathing dissipative solution can be regarded as the breathing kink-antikink molecule which has been observed in a modelocked fiber laser \cite{Nano}

From the expression \eqref{HPKM}, we know that the parameter $a_1$ determines the distance between two parallel HPKs of the molecule. If we take $a_1=1$, two HPKs are overlapped and the molecule is reduced to some kinds of central periodic kinks (CPK). Here, we list two special types of CPKs.\\
Case (a). Substituting
\begin{eqnarray}\label{P1}
&&\alpha=5,\ \beta=1, \ a_1=2,\ a_2=1,\ a_3=\frac{1}{10}, \ \kappa=-3
\end{eqnarray}
into \eqref{HPKM}, we obtain
\begin{eqnarray}\label{2.33}
u&=&-\frac{1}{15}\frac{10\exp \zeta+\cos\eta_1+45\sin\eta_1+10}
{10\cosh\zeta+\cos\eta_1+10},
\end{eqnarray}
where $\zeta=-\frac{2}{675}t-\frac{1}{15}x,\ \eta_1=\frac{676}5 t+3 x$.
Case (b). If the parameters of \eqref{HPKM} are given by
\begin{eqnarray}\label{P2}
&&\alpha=5,\ \beta=1,\  a_1=2,\ a_2=1,\ a_3=\frac{1}{10}, \ \kappa=-\frac34,
\end{eqnarray}
 we have $\zeta=-\frac{2}{675}t-\frac{1}{15}x,\ \eta_2=\frac{691}{320}t+\frac34 x$, then we obtain
\begin{eqnarray}\label{2.34}
u=-\frac{1}{60}\frac{40\exp\zeta+4\cos\eta_2+45\sin\eta_2+40}
{10\cosh\zeta+\cos\eta_2+10}.
\end{eqnarray}

\begin{figure}[htbp]
\centering
\subfigure(I){\includegraphics[height=5.5cm,width=6cm]{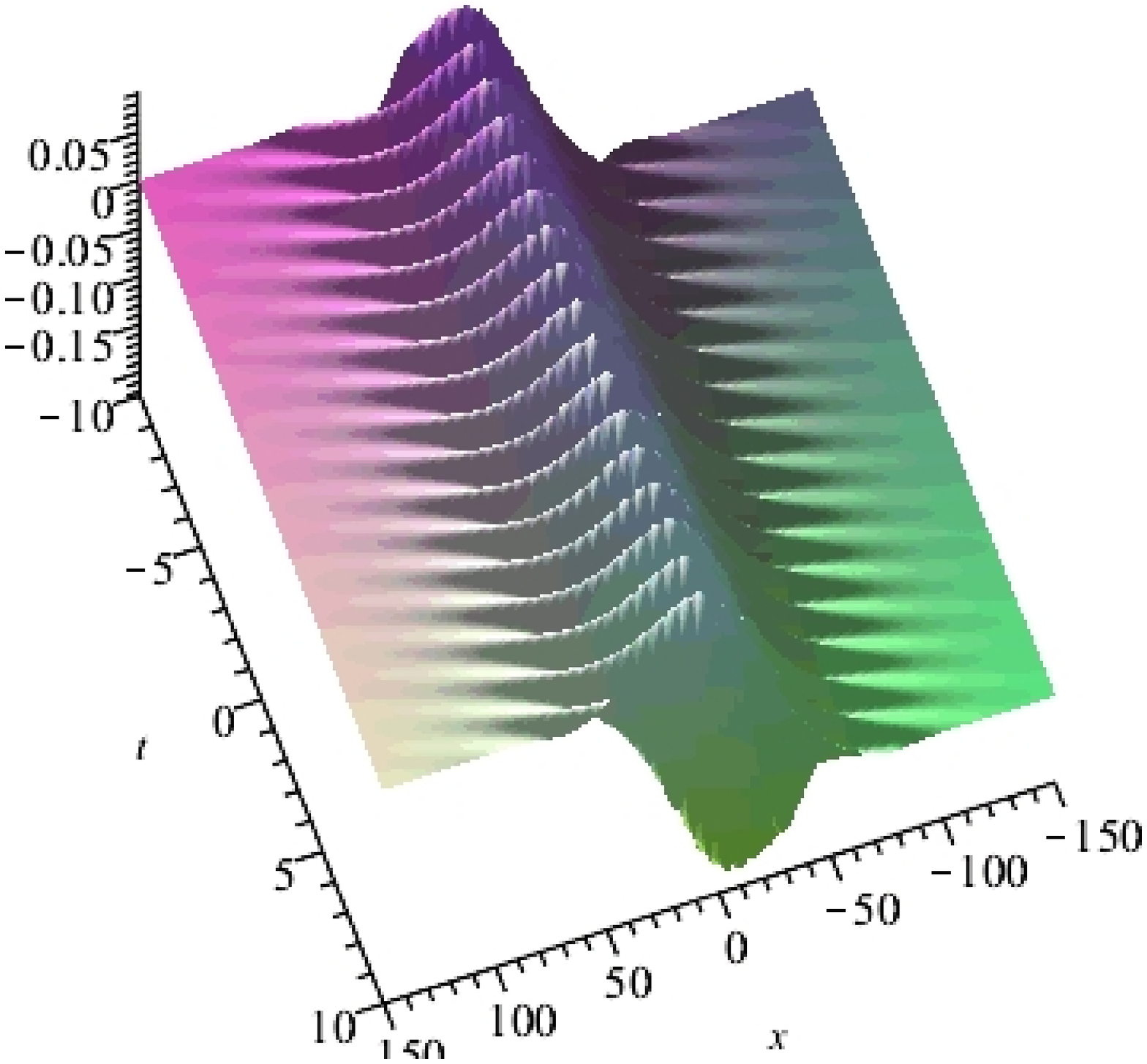}}
\subfigure(II){\includegraphics[height=5.5cm,width=6cm]{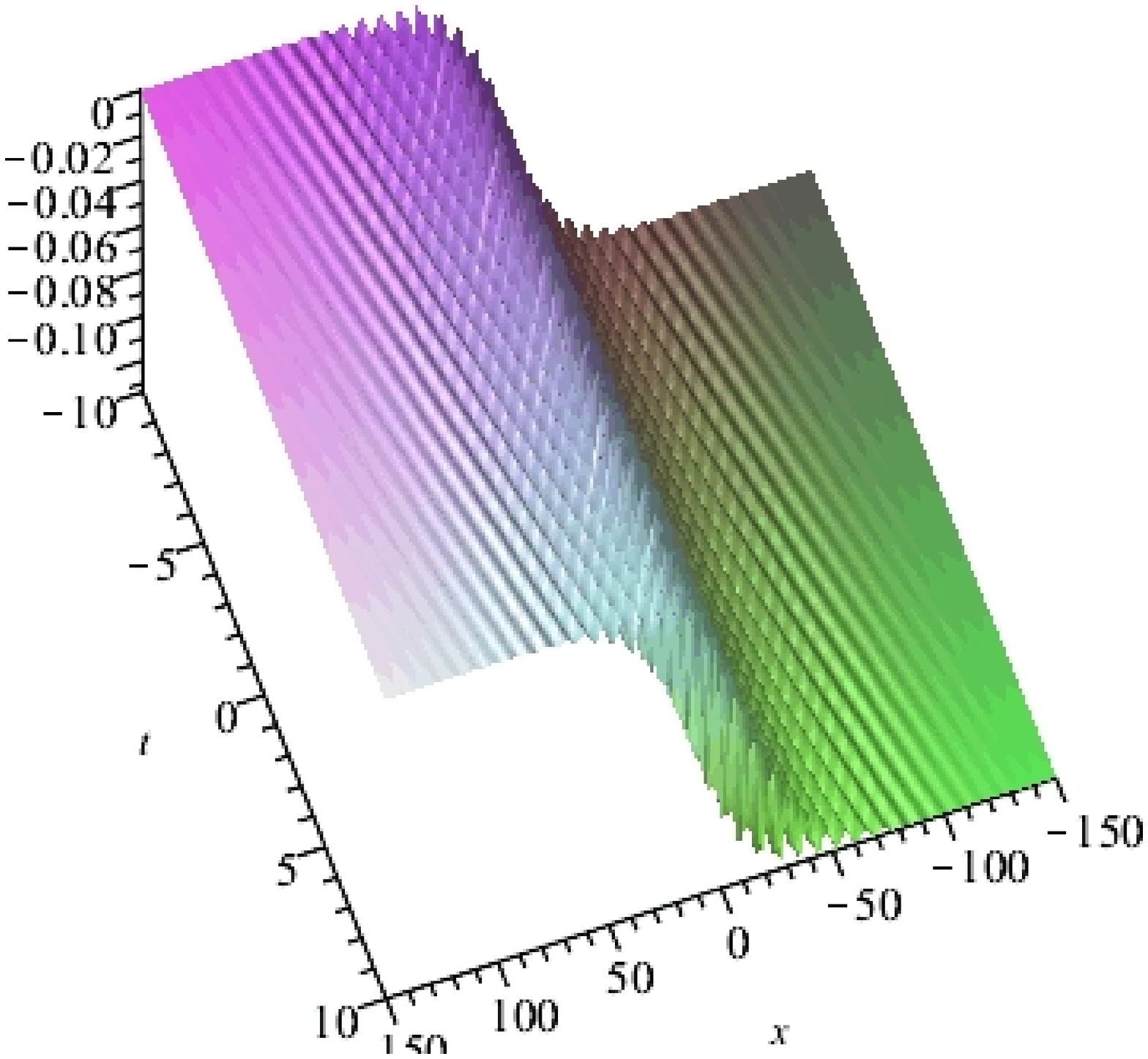}}
\caption{(I) The structure plot of the first type of CPK solution (\ref{2.33}); (II) The structure plot of the second type of CPK solution (\ref{2.34}). }
\label{CPK}
\end{figure}

The plot in \reffig{CPK} (I)  and (II) displays  the central periodic kink for (\ref{2.33}) and  (\ref{2.34}), respectively.  The central periodic kink (CPK) solution is composed of periodic solution in the center and a kink solution. Comparing  \reffig{CPK} (I)  with (II), we find the amplitude of Fig. (I) is higher than that of (II). From \reffig{HPK1}- \reffig{CPK}, we have  analyzed the fission and fusion between HPK and kink solutions, and the CPK solutions have also been presented. However, the Burgers equation and STO equations do not have such kind of solutions. It is verified  the STOB equation is a new systems possessing interesting and good structure and properties.

To find the fission and fusion properties related to HPK molecules, we take into account the solution of \eqref{N} for $N=5$ with one paired complex conjugate parameters.
Substituting
\begin{eqnarray}\label{para3}
&&\alpha=-1, \ \beta=1,\  k_1=\frac{2}{3},\ k_2=1,\ k_3=\frac13,\ k_4=k_3+\mbox{\rm i},\ k_5=k_3-\mbox{\rm i},\nonumber\\
&&a_1=\frac{1}{100},\ a_2=1, \ a_3=100,\ a_4=a_5=5, \ \xi_i=0, \ (i=1,\ \ldots,\ 5)
\end{eqnarray}
into \eqref{N} with $N=5$, we obtain
\begin{equation}\label{ab}
u=\frac23\frac{\mbox{\rm e}^{2X}+150\mbox{\rm e}^{x}+500(10-3\sin Y+\cos Y)\mbox{\rm e}^{X}}
{\mbox{\rm e}^{2X}+100(1+\mbox{\rm e}^{x})+1000(10+\cos Y)\mbox{\rm e}^{X}},
\end{equation}
where $X=\frac13x-\frac2{27}t,\ Y=x-\frac43t$.

\begin{figure}[htbp]
\centering
\subfigure(I){\includegraphics[height=5.5cm,width=6cm]{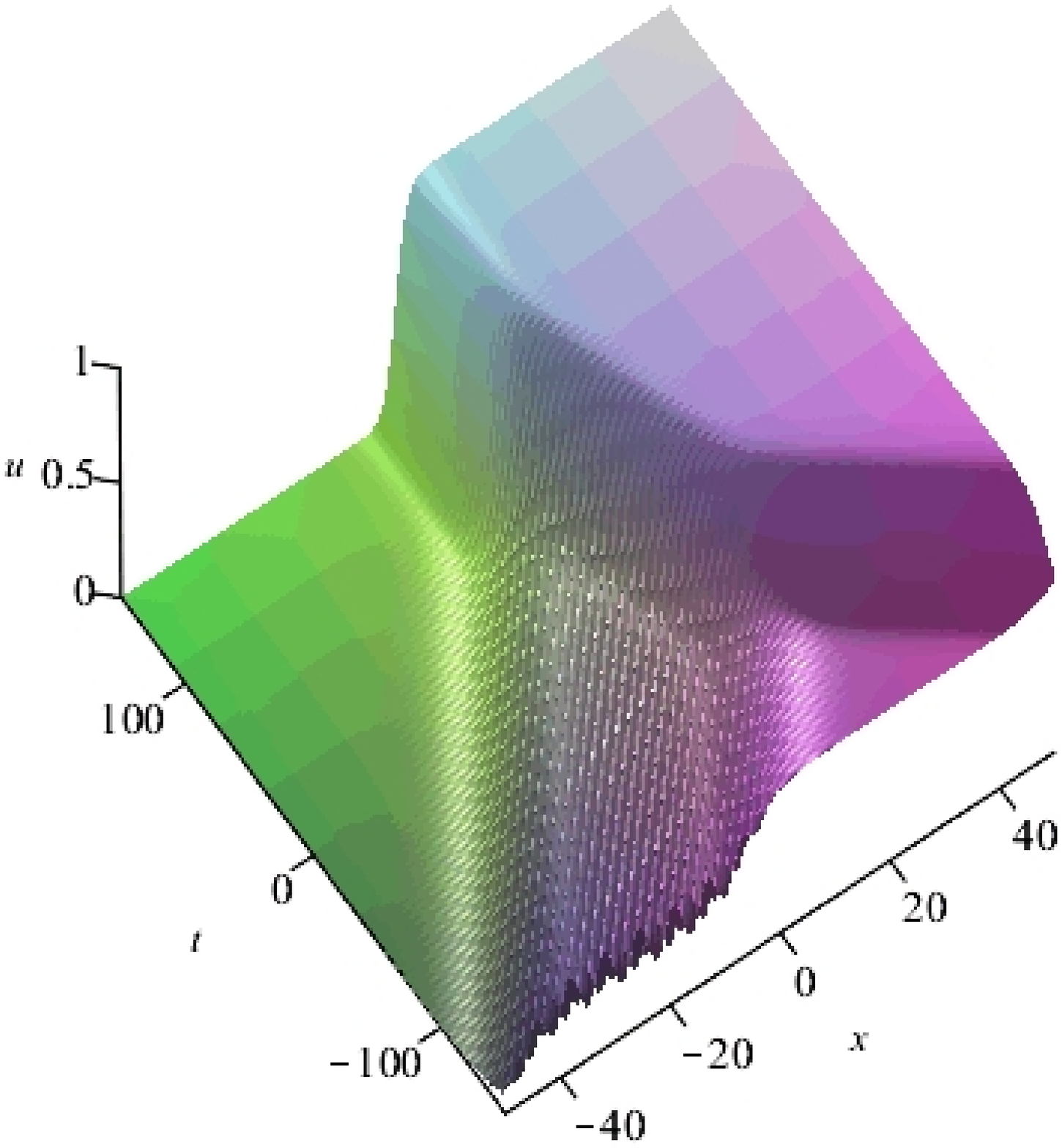}}
\subfigure(II){\includegraphics[height=5.5cm,width=6cm]{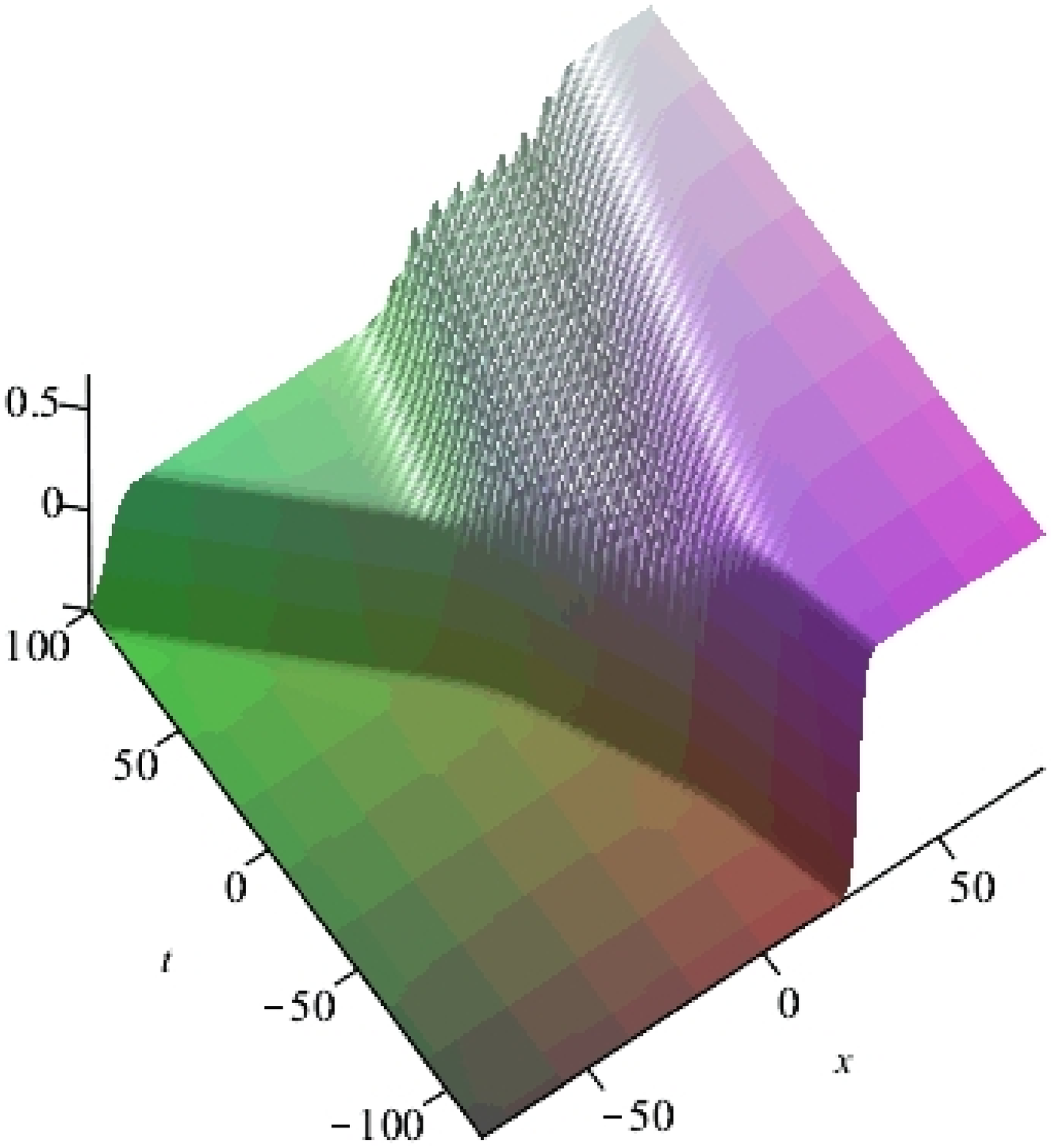}}
\caption{(I) Absorption interaction expressed by \eqref{ab}: An HPK molecule is absorbed by a usual kink, (II) Escape interaction described by \eqref{2.53}: An HPK molecule is split out from the usual kink.}
\label{SM8}
\end{figure}

\reffig{SM8} (I) reveals an HPK molecule and a usual kink are fused to an ordinary kink. In other words, a usual kink absorbs an HPK molecule. In the same way, an HPK molecule may be split out from a usual kink. By substituting
\begin{eqnarray}\label{para4}
&&\alpha=-1, \beta=1, k_1=\frac{2}{3},\ k_2=-\frac12,\ k_3=\frac13,\ k_4=k_3+\mbox{\rm i},\ k_5=k_3-\mbox{\rm i},\nonumber\\
&&a_1=\frac{1}{100},\ a_2=1, \ a_3=100,\ a_4=a_5=5, \ \xi_i=0, \ (i=1,\ \ldots,\ 5),
\end{eqnarray}
into \eqref{N} with $N=5$, we get
\begin{eqnarray}\label{2.53}
u=\frac23\frac{\mbox{\rm e}^{2X}-75\mbox{\rm e}^{-Z}+500(10-3\sin Y+\cos Y)\mbox{\rm e}^{X}}
{\mbox{\rm e}^{2X}+100(1+\mbox{\rm e}^{-Z})+1000(10+\cos Y)\mbox{\rm e}^{X}},
\end{eqnarray}
where $X=\frac13x-\frac2{27}t,\ Y=x-\frac43t,\ Z=\frac12x+\frac38t$.

\reffig{SM8} (II) indicates that a usual kink is split into an ordinary  kink and an HPK molecule. In alternatively speaking, an HPK molecule is escaped from a usual kink.

In this section, we  investigate the soliton (kink) molecules, HPK molecules, breathing dissipative solitons (breathing kink-antikink molecules) and the fissions and fusions between HPK and kink molecules from the multiple solitary wave solutions of the STOB equation by introducing velocity resonant mechanism.

\section{Summary}

By means of  introducing the velocity resonant mechanism,  we investigate the solitary wave solutions including soliton molecules, HPKs, HPK molecules and CPKs for the STOB equation. The fission and fusion phenomena have been analyzed not only among (kink) solitons but also among usual kinks, HPKs and HPK molecules. These  resonant solitary waves have changed the corresponding amplitude, widths, velocity and wave shape after the interactions. It is proved that the Bugers equation only admit fusion phenomenon. However, the STOB equation possess both fission and fusion with imposing  the resonant conditions. Furthermore, we establish some new types of solutions and molecules in STOB equation which Burgers and STO equation do not have. The fission and fusion of HPK and kink solutions of multiple solitary solutions have been discussed and CPK solutions have also been derived. Meanwhile, we investigate many kinds of molecules for STOB equation, such as soliton molecules, kink  molecules, HPK molecules and  the breathing soliton molecules.

The kink-kink molecules as shown in Fig. 6 II may be used to describe the well known layer dislocations in solid-state physics physics and the domain walls known in magnetic materials science.
The breathing dissipative solitons (breathing kink-antikink molecule) as shown in Fig. 10 may be one of the candidate to describe the known observations in experiments \cite{SA}.
How to discuss the structure and properties of  molecules  for other integrable systems  deserves further study. The various types of molecules presented in this article  should provide new insight into applications in  physics.

\section{Acknowledgements}
The work was sponsored by the National Natural Science Foundations of China (Nos. 11435005, 11975131, 11965014 and 11605096) and K. C. Wong Magna Fund in Ningbo University.

\end{document}